\documentclass[epsfig,useAMS,useAMSmath,usenatbib]{mn2e}

\usepackage{url}

\usepackage{graphics}
\usepackage[dvips]{graphicx}
\usepackage {layout}

\newcommand{\Msun}{M_\odot}

\newcommand{\beq}{\begin{equation}}
\newcommand{\eeq}{\end{equation}}

\newcommand\aj{AJ}
\newcommand\apj{ApJ}

\newcommand\apjs{ApJS}
\newcommand\mnras{MNRAS}

\newcommand\nat{Nature}
\newcommand\aap{A\&A}

\title[Recovering cores and cusps in simulations]{Recovering cores and
cusps in dark matter haloes using mock velocity field observations}

\author[Kuzio de Naray \& Kaufmann]
{Rachel Kuzio de Naray$^1$\thanks{Current Address: Department of Physics, 
Royal Military College of Canada, P.O. Box 17000, Station Forces, Kingston, 
ON, K7K 7B4, Canada}\thanks{E-mail: kuzio@rmc.ca} 
and Tobias Kaufmann$^{2}$\thanks{E-mail: tobias.kaufmann@phys.ethz.ch}\\
$^1$Center for Cosmology, Department of Physics and Astronomy, University 
of California, Irvine, CA 92697-4575 \\$^2$ Institute of Astronomy, 
ETH Z\"urich-H\"onggerberg, CH-8093 Z\"urich, Switzerland}

\begin{document}

\pagerange{\pageref{firstpage}--\pageref{lastpage}} \pubyear{} 

\maketitle

\begin{abstract}
  We present mock DensePak Integral Field Unit (IFU) velocity fields,
  rotation curves, and halo fits for disc galaxies formed in spherical
  and triaxial cuspy dark matter haloes, and spherical cored dark
  matter haloes.  The simulated galaxies are ``observed'' under a
  variety of realistic conditions to determine how well the underlying
  dark matter halo can be recovered and to test the hypothesis that
  cuspy haloes can be mistaken for cored haloes.  We find that the
  appearance of the velocity field is distinctly different depending
  on the underlying halo type.  We also find that we can successfully
  recover the parameters of the underlying dark matter halo.  Cuspy
  haloes appear cuspy in the data and cored haloes appear cored.  
    Our results suggest that the cores observed using high-resolution
    velocity fields in real dark matter-dominated galaxies are genuine
    and cannot be ascribed to systematic errors, halo triaxiality, or
    non-circular motions.
\end{abstract}
\section{Introduction}

Cold Dark Matter (CDM) is one of the main constituents of the current
standard model in cosmology, $\Lambda$CDM.  N-body simulations based
on the CDM paradigm predict dark matter haloes best-described by a
steep (``cuspy'', ``NFW'') power law mass density distribution
\citep[e.g.][]{Navarro96a}. While the exact slope of the inner density
distribution remains under debate
\citep[e.g.][]{Moore99,Graham06,Stadel09,Navarro10}, simulations all
agree on producing a cuspy profile with a slope of $\alpha \sim -1$ in
the central part of the halo \citep[see the review of][]{deBlok10}.

Observed rotation curves of galaxies typically need a massive dark
matter halo with a nearly constant density core in order to describe
the data well. Ideal objects for observationally studying the
``cusp-core problem'' are low surface brightness (LSB) galaxies which
are late-type, gas-rich, dark matter-dominated disc
galaxies. Observations using long-slit spectra in H$\alpha$
\citep[e.g.][]{deBlok02} or high-resolution two-dimensional velocity
fields \citep[][hereafter K06 and K08, respectively]{Kuzio06,Kuzio08}
show that a cored dark matter distribution (described e.g.~by an
pseudoisothermal sphere profile) provide a better fit to the data than
a cuspy NFW dark matter profile.  In the few cases where NFW models
  do reasonably fit the rotation curve data, the predicted
  values for concentrations $c_{200}$ are very low and the circular
  velocity $V_{200}$ at the virial radius very high, inconsistent with
  the cosmological $(c_{200},V_{200})-relation$, \citep[see the review
  of][]{deBlok10}.

Most observational problems such as
pointing errors, centering offsets and non-circular motions are likely
too small to cause cusps to be mistaken for cores in the observations
\citep{deBlok10}, but halo triaxiality has been claimed to explain the
presence of cores.  The simulation works of \citet{Hayashi07} and
\citet{Bailin07}, for example, showed that due to systematic
non-circular motions introduced by a triaxal halo, cusps could be
interpreted for cores if long-slit rotation curves were
used. Similarly, bars inducing non-circular motions combined with
projection effects can create the illusion of a constant-density core
in a circular velocity analysis \citep []{Valenzuela07}. Pressure
support in the rotating gas of low mass galaxies could mask the
underlying dark matter distribution, but the importance of this effect
remains uncertain \citep []{Dalcanton10}.

\defcitealias{Kuzio06}{K06}
\defcitealias{Kuzio08}{K08}

 The aim of this work is to investigate whether cusps in (LSB) dark
  matter haloes could still be mistaken for cores if high-resolution
  two-dimensional velocity fields data are available  \citep[as
e.g.~in][]{Kuzio09}. In order to do so, we use self-consistent
high-resolution simulations of (LSB) galaxy formation and ``observe''
the simulated galaxies using the DensePak Integral Field Unit. We
compare the mock velocity fields for different signatures from cuspy
and cored haloes, spherical and triaxial potentials, and the effect of
supernovae feedback.  We also determine how well the underlying dark
matter halo potential can be recovered.

The outline of the paper is as follows. In Section 2, we present the
modeling of our initial conditions, the code used for the time
evolution and the halo parameters derived from the simulations. In
Section 3, we discuss the process of observing the simulations.  The
mock velocity fields and rotation curves are presented in Section 4.
In Section 5, we present the halo fits to the mock data.  A discussion
and summary are presented in Section 6.
\section{Smoothed particle hydrodynamic simulations}

We perform non-cosmological high-resolution N-Body/SPH (smoothed
particle hydrodynamics) simulations of disc galaxy formation by
cooling a rotating gaseous mass distribution inside equilibrium cuspy
spherical, cored spherical, and cuspy triaxial dark matter haloes
similar to what is described in \citet{Kaufmann07a}. The halo
parameters were chosen to form extendend discs with masses of a few
$\times 10^8 \Msun$ without bars or massive bulges in a dark halo with
$c_{200} \sim 8.5$, as is common in low mass galaxies
\citepalias{Kuzio06}.  While these simulations are not modeling the
full complexity of disc formation (e.g. mergers are missing) the
dynamics of a disc dominated by a dark halo (i.e. the late phase of
disc formation) is followed self-consistently and with high numerical
resolution.
\subsection{Initial conditions}

Our initial conditions for the spherical runs comprise an isolated
equilibrium cuspy or cored galaxy-sized halo with an embedded spinning
gaseous mass distribution initially in hydrostatic equilibrium. The
halo models are built as in \citet{Kazantzidis04}, and hence they
include a self-consistent description of the velocity distribution
function.  Those models with virial mass of $M_{200} = 1.3 \times
10^{11} \Msun$ were constructed using the $\alpha\beta\gamma$ models
as described in \citet{Zemp08}. Cuspy haloes were modeled using the
NFW form ($1, 3,1$ for $\alpha,\beta,\gamma$) and $c_{200} = 8.5$
following \citet{Maccio08}. For cored haloes, the parameters were set
to $0.8,3,0$ and $c_{200} = 24$.   This choice produced an initial core with a 
radius of $\sim$ 2 kpc,  similar to what has been found in observational work \citepalias{Kuzio06}.

We initialise a fraction of the virial halo mass, $f_{\rm b}=0.01$, as
a hot baryonic component with the same radial distribution as the dark
matter.  The initial temperature structure of the gaseous halo is
calculated by solving the equation for the hydrostatic balance of an
ideal gas inside a dark matter halo. The gas halo is spun with a gas
spin parameter of $\lambda_g = 0.06$, and the dark matter particles
are initialised with no net angular momentum (see \citet{Kaufmann07a}
for details). We note that the baryon fraction is much lower than the
universal baryon fraction of $\sim 0.17$, though this is not
unexpected in low mass galaxies \citep{McGaugh10}.  This choice of
$f_{\rm b}$ forms discs with masses of a few $\times 10^8 \Msun$ in
our models.

The models use $2.5\times10^{6}$ dark matter and $2\times10^{5}$ gas
particles (mass of a dark particle $\sim 7\times10^{4} \Msun$) and
have a spatial resolution (softening length) of $100$ pc.

\begin{figure}
\includegraphics[scale=0.4]{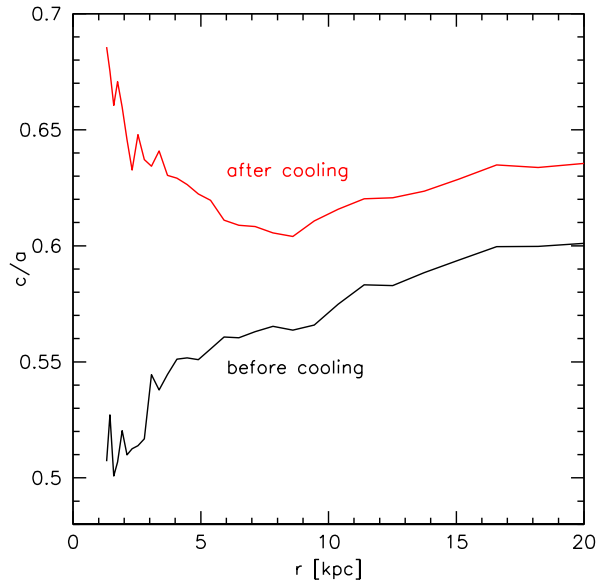}
\caption{The ratio between the short and long axes of the dark matter
density distribution of the triaxial halo before cooling (black line)
and at the time of the disc analysis (red line).  \label{axisratio}}
\end{figure}

To construct an equilibrium triaxial halo in which the gas acquires
its angular momentum through an equal mass merger
\citep[see][]{Moore04,Kaufmann07a}, we start with two spherical,
non-rotating NFW haloes each with half of the mass and particles of
the fiducial model considered previously. The two haloes are placed at
a separation of twice their virial radius. One of them is given a
transverse velocity of $27$ km s$^{-1}$, and the haloes are allowed to
merge. This net velocity (determined using a trial-and-error
procedure) results in a final gas distribution that has a similar spin
parameter to that used in the spherical models.  The triaxial halo
resulting after the merger is first evolved for $15$ Gyr with an
adiabatic equation of state for the gas. The inner part of the dark
halo is significantly triaxial, $c/a\approx0.6$ (and $b/a\approx0.7$)
and the subsequent formation of the disc due to cooling modifies the
shape of the dark matter only weakly (see Figure \ref{axisratio} and
also \citet{Kazantzidis10}).

\subsection{Time evolution and derived halo parameters}

All of the systems were then evolved for $\sim 4.5$ Gyr using the
parallel TreeSPH code \textsc{Gasoline} \citep{Wadsley04}, which is an
extension of the pure N-Body gravity code \textsc{Pkdgrav} developed
by \citet{Stadel01}. The code implements Compton and radiative
cooling, as well as the star formation and ``blastwave feedback''
prescriptions detailed in \citet{Katz96} and \citet{Stinson06}. In
order for star particles to form, the gas temperature must be
$<30,000$ K, the local gas density must be $>0.1$ cm$^{-3}$ and other
local gas particles must define a converging flow.  Stars then form
according to the Schmidt law \citep{Kennicutt98} using a
\citet{Miller79} initial mass function. The supernova (SN) feedback
model creates turbulent motions in nearby gas particles that keep them
from cooling and forming stars. Because of the lack of metal and
molecular cooling, the efficiency of our cooling function drops
rapidly below $10^4$ K acting as a temperature floor similar to what
has been used in \citet{Kaufmann07b}. Star formation is only switched
on in the cuspy model using the feedback prescription\footnote{Heating
by a uniform UV background from QSO following \citet{Haardt96} has
also been used.}; the other simulations are cooling-only runs.

\begin{table}
\begin{center}
  \caption{Dark halo parameters derived from the simulations after the
    evolution with cooling\label{parameters}}
\begin{tabular}{|lcc|}
\hline  
\hline
Name   & $R_{200}$ or $r_{core}$ &  $c_{200}$ or $\rho_{core}$\\
\hline
Cuspy & 104.8 kpc & 8.4\\
Cuspy SN & 104.3 kpc & 8.3\\
Triaxial & 102.9 kpc & 8.8\\
Cored & 1.5 kpc & $69\times 10^{-3} \Msun$ pc$^{-3}$\\
\hline
\end{tabular}
\end{center}
\end{table}

At the time of the analysis, thin, extended baryonic discs resembling
LSB galaxies have formed in the centre of the haloes.   This
  redistribution of the baryonic matter due to cooling also affects
  the underlying dark matter distribution, see below. The
discs do not show bars or massive bulges, have baryonic masses of
$\sim 5\times 10^{8} \Msun$, and star formation rates, $SFR \sim
0.33\Msun$ yr$^{-1}$, comparable to those observed in LSB galaxies
\citep[]{Wyder09}.   The discs are dark matter-dominated with the 
  baryonic mass reaching up to $30\%$ of the total mass within
  the inner $0.5$ kpc and far less beyond.  The galaxies formed in the
 spherical simulations lie between the mean and the
upper $2\sigma$ deviation of the (observational) Radius-Velocity
relation presented in \citet{Dutton07}, while the galaxy formed in the
triaxial simulation lies just above the upper $2\sigma$ deviation from
the mean.  These somewhat large radial scale-lengths are not
unexpected as the spin parameter chosen in the initial conditions was
above the mean.  Additionally, it has been suggested by
\citet{Maccio07} that LSB galaxies reside in haloes with spins higher
than average.

 We derive halo parameters after the evolution with cooling directly
from the simulations and list them in Table \ref{parameters}. 
  Additionally, in Table \ref{parameters} we establish our naming
  convention for the simulations.  We refer to the spherical cuspy
  simulation without feedback as `Cuspy'; the spherical cuspy
  simulation with feedback is referred to as `Cuspy SN'.  The triaxial
  cuspy simulation is referred to as `Triaxial' and the spherical
  cored simulation is referred to as `Cored'.  Motivated by the
  parameterisations used in observational works
  \citep[e.g.][]{deBlok10} we fit an NFW profile \citep{Navarro96a} to
the dark matter density curves derived from the cuspy simulations and
a pseudoisothermal sphere profile (as in e.g.~\citetalias{Kuzio06}) to
the cored simulation. 

The initial and final  halo shapes do not
differ significantly, even in the simulation with feedback.  In our
simulations, massive dark matter haloes are in place at the beginning
of the simulation and the low mass present in baryons
($f_{baryons}=0.01$) is  of little importance for the dynamics of
  the dark matter (specifically, the mass of blown-out baryons is too
  small to affect the dynamics of the dark halo). The only potential
influence from the supernova feedback is a disturbance of the velocity
field of the gas in the galactic disc.  This is different than the
cosmological simulations of \citet{Governato10} (using the same
feedback prescription but a higher minimal density for star formation)
who find that cored dark matter haloes are created due to strong
outflows from supernovae in dwarf galaxies.  The outflows remove the
central dark matter cusp at high redshift when the dark matter halo is
less massive.

\section{``Observing'' the Simulations}

\begin{figure}
\begin{center}
\includegraphics[scale=0.5]{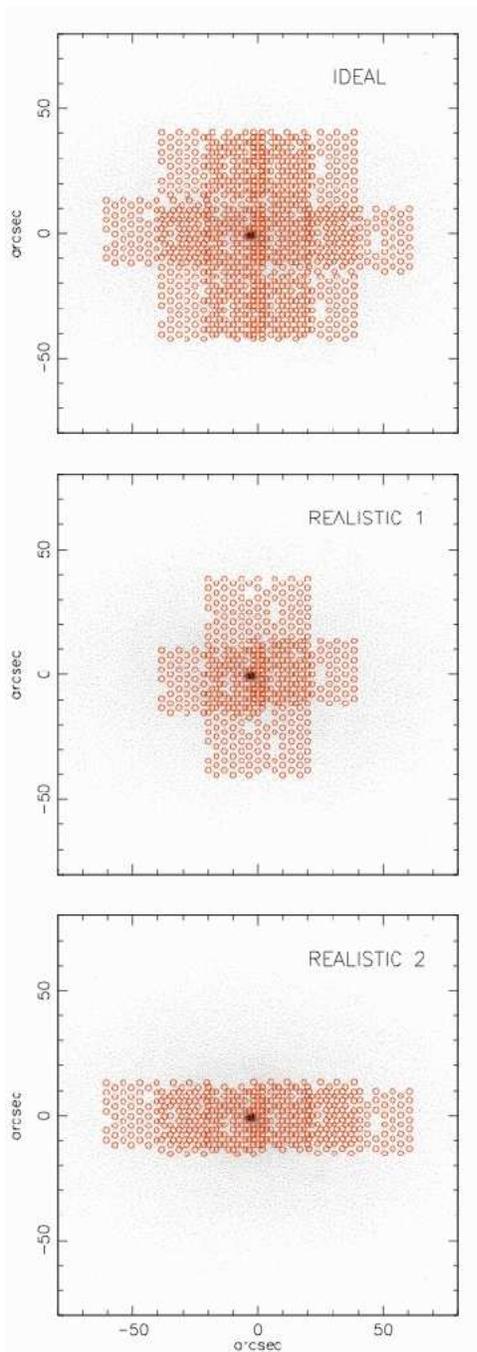}
\caption{Configurations of the IFU fibers (red circles) shown on the
galaxy (grey points).  \textit{Top:} There is maximum spatial coverage
across the galaxy in the Ideal configuration and all fibers provide a
velocity measurement. \textit{Middle:} In the Realistic 1
configuration, the reduced number of fibers cover the central part of
the galaxy.   Additionally, $\sim$40\% of the fibers shown are randomly turned off so that not all fibers
provide a velocity measurement.  \textit{Bottom:} The fibers in the
Realistic 2 configuration are along the galaxy major axis.  As in the
Realistic 1 configuration, not all fibers provide a velocity
measurement.  \label{configurations}}
\end{center}
\end{figure}

Each galaxy+halo simulation is ``observed'' under a variety of
instrumental configurations and physical properties designed to
represent a range of mock data quality from ideal to realistic.  We
test three distances, 15~Mpc, 30~Mpc, and 45~Mpc, corresponding to
spatial resolutions of $\sim$ 0.07~kpc/$\arcsec$, 0.15~kpc/$\arcsec$,
and 0.22~kpc/$\arcsec$, respectively.  The choice of these spatial
resolutions was motivated by the spatial resolution obtained for the
LSB galaxies observed in \citetalias{Kuzio06} and
\citetalias{Kuzio08}, as well as a requirement that a 3$\arcsec$ IFU
fiber provide sub-kiloparsec resolution.  We test three galaxy
inclinations, $i$ = 30$\degr$, 50$\degr$, 65$\degr$, the range again
being motivated by the data presented in \citetalias{Kuzio06} and
\citetalias{Kuzio08}.    There is an inherent trade-off between inclination and obtaining 
a well-sampled velocity field, and most observational studies target galaxies having 
inclinations within, or very close to, this range \citep[e.g.][]{Spano08,Blais04}. Not all lines of sight in nonaxisymmetric 
systems are equivalent \citep{Kuzio09}, so we also rotate the
galaxy+halo system by a random angle in the plane of the disc.  For
consistency, we do this for the spherical simulations in addition to the
triaxial simulation.

DensePak, an integral field spectrograph on the 3.5-m WIYN telescope
at the Kitt Peak National Observatory (KPNO), is used to ``observe''
each simulated galaxy.  DensePak is a 43$\arcsec$~$\times$~28$\arcsec$
fixed array of 3$\arcsec$ fibers with 3.84$\arcsec$ separations.  We
model the 85 working, and five missing or broken, fibers in the main
bundle.  We test three configurations of DensePak fiber alignment (see
Figure \ref{configurations}).  In the Ideal case, 11 pointings of the IFU
array provide maximum spatial coverage across the galaxy.  In the
Realistic 1 and Realistic 2 cases, there are 5 pointings of the IFU
array.  In the Realistic 1 case, the IFU fibers cover the central
region of the mock galaxy.  The IFU fibers are along the mock galaxy
major axis in the Realistic 2 configuration.  The velocity recorded
for each fiber is the average of all the line-of-sight velocities of
the gas particles falling within the fiber.  In the Ideal case, all
935 IFU fibers provide a velocity measurement.  To simulate the sparse
H$\alpha$ emission observed in some real galaxies, $\sim$~40\% of the
fibers in the Realistic 1 and 2 cases are randomly turned off so that
no velocity measurement is obtained.  Finally, we ensure that the
fiber-to-fiber velocity variation is 6-10~km~s$^{-1}$, consistent with
the velocity dispersion measured in the \citetalias{Kuzio06} galaxies.

Mock velocity fields are obtained using all three IFU fiber
configurations at the high and medium spatial resolutions.  Only the
Ideal and Realistic 2 configurations are used at low spatial
resolution because of the reduced angular size of the galaxy.  The
three galaxy inclinations are tested at each spatial resolution and
IFU fiber configuration.  For each galaxy+halo simulation, we obtain
five mock velocity fields at each combination of galaxy parameters and
IFU configurations.  Our final data set includes 120 mock velocity
fields per galaxy+halo simulation, for a total of 480 mock velocity
fields.
\section{Mock Data}

In this section, we present examples of the mock velocity fields and
derived rotation curves for each of the four galaxy+halo simulations.
\subsection{Mock velocity fields}
\label{MockVFsection}

\begin{figure}
\begin{center}
\includegraphics[scale=0.5]{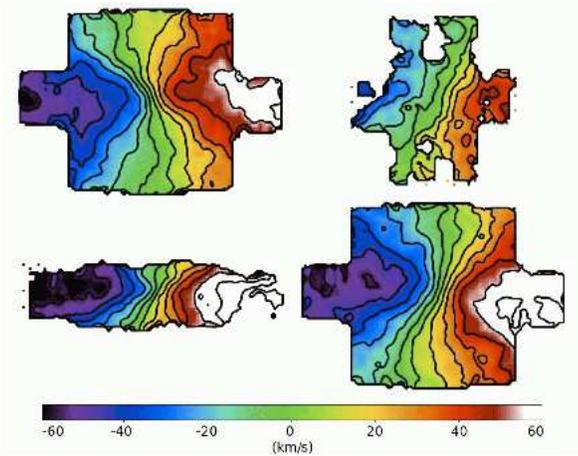}
\caption{Sample mock velocity fields of the triaxial  cuspy simulations;
isovelocity contours are at 10 km s$^{-1}$ intervals.  The velocity
fields in the upper left and lower right are examples of the Ideal
configuration; the upper right and lower left are examples of the
Realistic 1 and Realistic 2 configurations, respectively.  The
kinematic and photometric major axes are not aligned, the kinematic
major and minor axes are not perpendicular, and the minor axis is
often twisted.   \label{triaxialvfs}}
\end{center}
\end{figure}

\begin{figure}
\begin{center}
\includegraphics[scale=0.5]{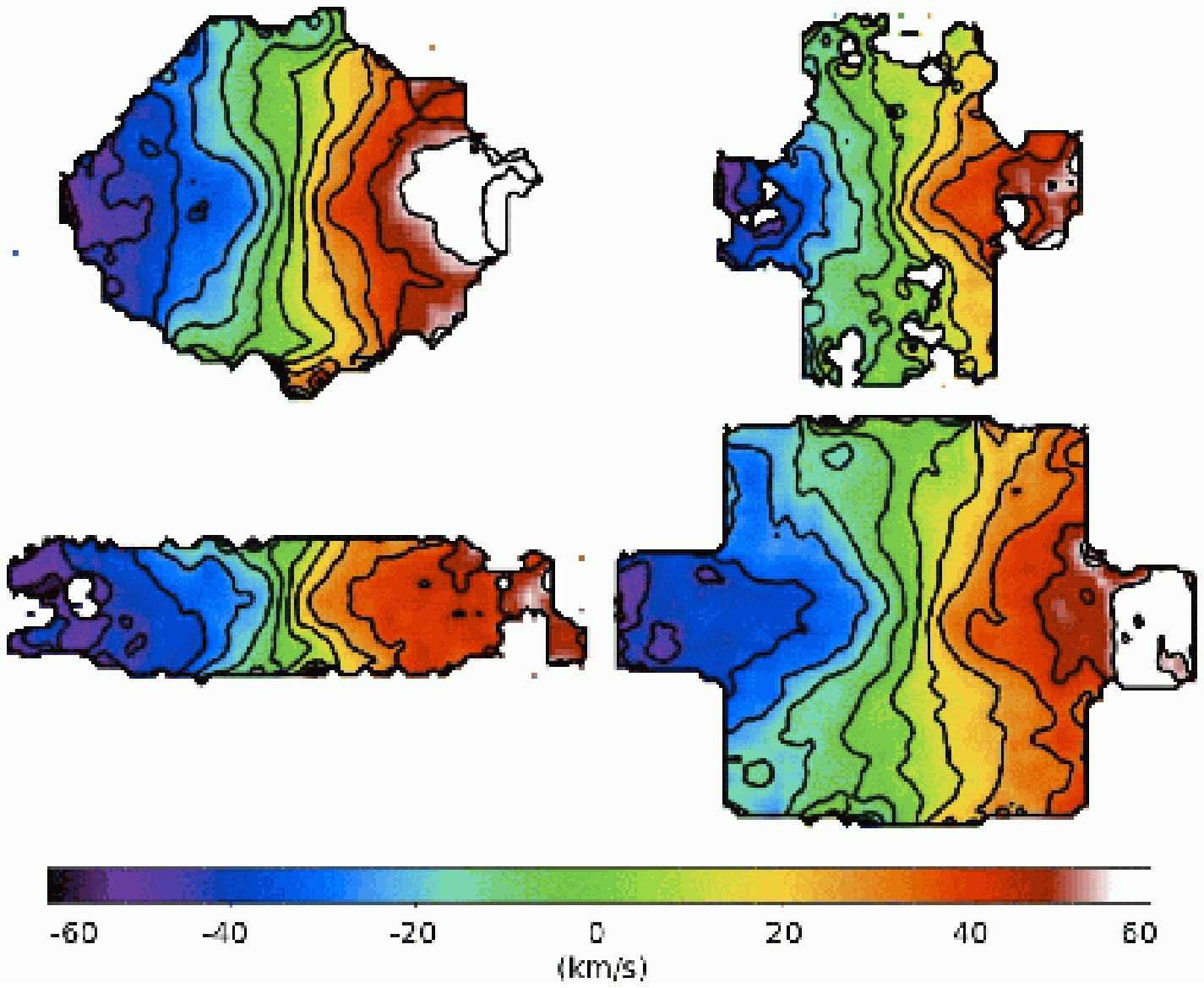}
\caption{Sample mock velocity fields of the spherical cuspy
simulations; isovelocity contours are at 10 km s$^{-1}$ intervals.
The layout is the same as Figure \ref{triaxialvfs}.  The kinematic and
photometric major axes are aligned, the kinematic major and minor axes
are perpendicular, and the isovelocity contours are pinched at the
centre of the velocity field.   \label{cuspyvfs}}
\end{center}
\end{figure}

\begin{figure}
\begin{center}
\includegraphics[scale=0.5]{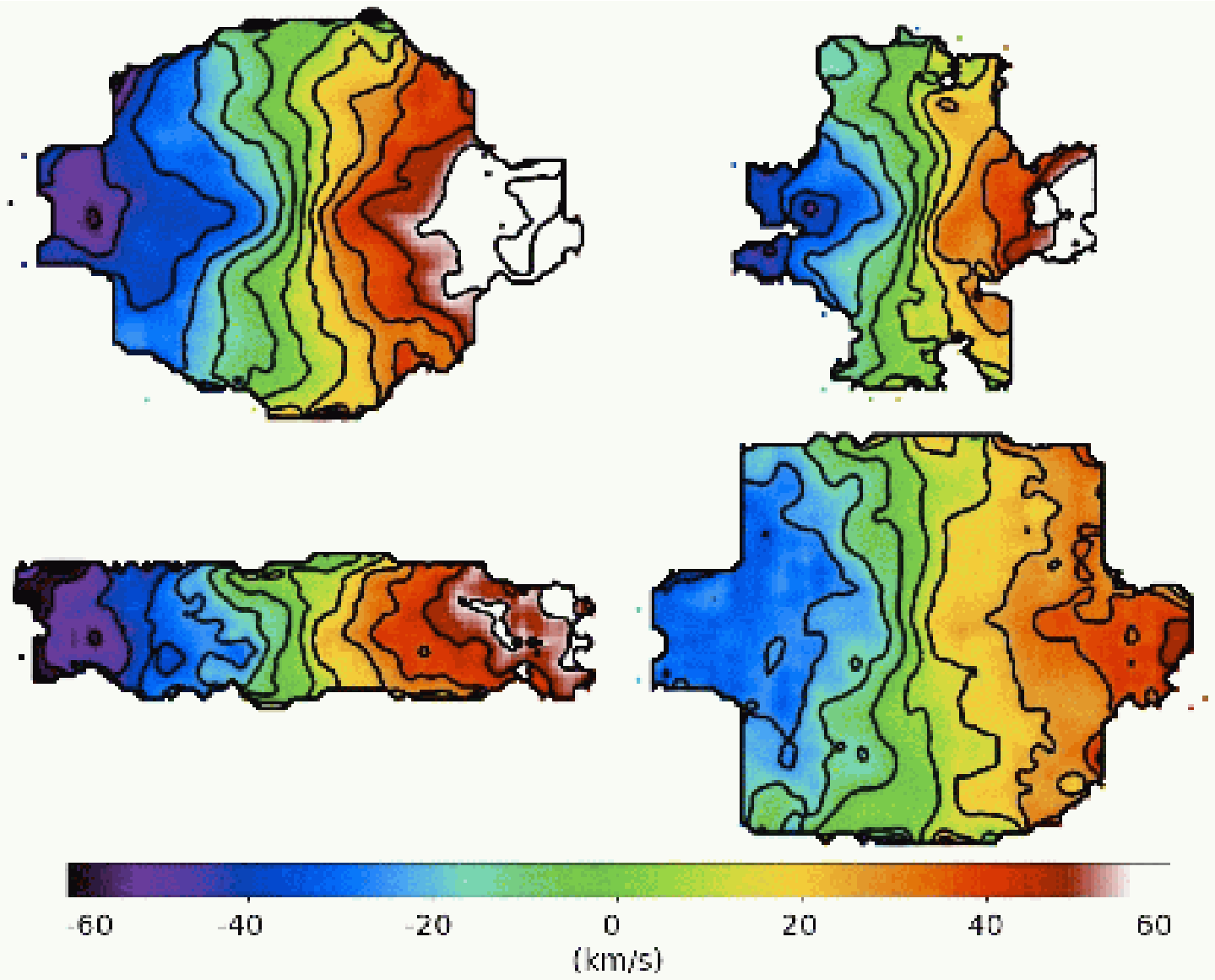}
\caption{Sample mock velocity fields of the spherical cuspy SN
simulations; isovelocity contours are at 10 km s$^{-1}$ intervals.
The layout is the same as Figure \ref{triaxialvfs}.  The kinematic and
photometric major axes are aligned, the kinematic major and minor axes
are perpendicular, and the isovelocity contours are pinched at the
centre of the velocity field.   \label{cuspySNvfs}}
\end{center}
\end{figure}

\begin{figure}
\begin{center}
\includegraphics[scale=0.5]{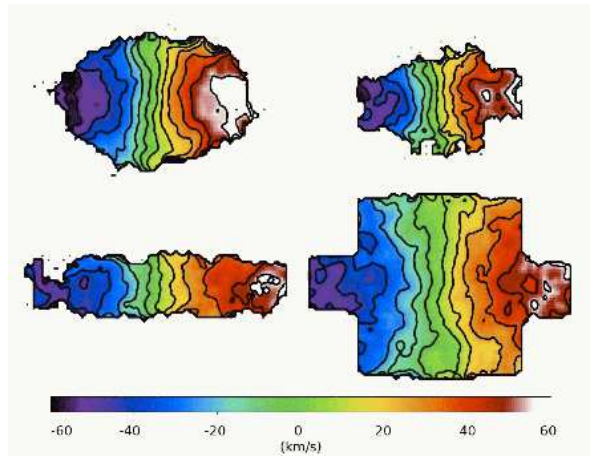}
\caption{Sample mock velocity fields of the spherical cored simulations;
isovelocity contours are at 10 km s$^{-1}$ intervals.  The layout is
the same as Figure \ref{triaxialvfs}.  The kinematic and photometric
major axes are aligned, the kinematic major and minor axes are
perpendicular, and the isovelocity contours are parallel at the centre
of the velocity field.   \label{coredvfs}}
\end{center}
\end{figure}

In Figures \ref{triaxialvfs}-\ref{coredvfs}, we show four
representative examples of the mock velocity fields obtained for each
simulation, covering the range of spatial resolutions, inclinations,
and IFU configurations.  The velocity fields in the upper left and
lower right are examples of the data obtained using the Ideal
configuration while the upper right shows an example of the Realistic
1 configuration and the lower left an example of the Realistic 2
configuration. Figure \ref{triaxialvfs} contains mock velocity fields
for the triaxial simulation.  Mock velocity fields for the spherical
cuspy simulation are in Figure \ref{cuspyvfs}, those including
feedback are in Figure \ref{cuspySNvfs}.  Figure \ref{coredvfs}
contains mock velocity fields for the cored simulation.

\begin{figure}
\begin{center}
\includegraphics[scale=0.4]{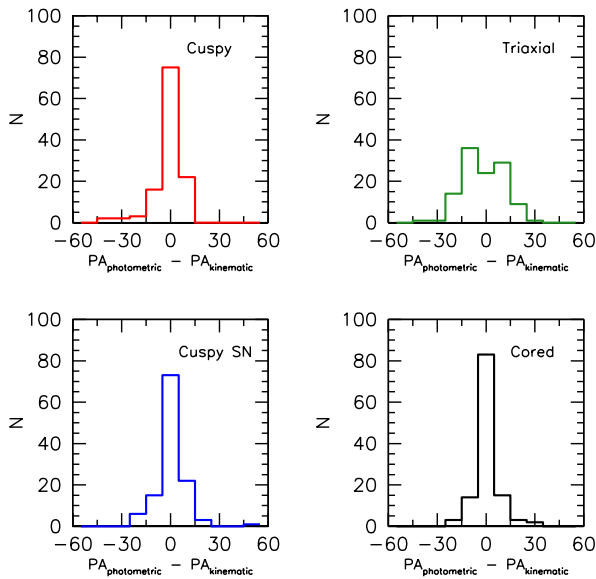}
\caption{Difference in degrees between the photometric and kinematic
major axes as measured at a radius of 19.5$\arcsec$, approximately
half the length of a single DensePak pointing. The axes are aligned in
the majority of the spherical cuspy (with and without feedback) and
cored mock velocity fields; the axes are frequently misaligned in the
triaxial mock velocity fields.    \label{pahist}}
\end{center}
\end{figure}

In general, the mock velocity fields of the triaxial simulation
(Figure \ref{triaxialvfs}) are characterised by misaligned kinematic
and photometric major axes, and non-perpendicular kinematic major and
minor axes (see Figure \ref{pahist}). Most noticeably, the minor
axis is often twisted.  Though no physical bar is present in the
simulation, these are features commonly observed in the velocity
fields of barred galaxies
\citep[e.g.][]{Bosma81,Hernandez05}. Bar-like dynamical signatures
(without an obvious bar) are also expected based on simulation work
for subdominant discs in triaxial halos \citep[]{Kazantzidis10}. The
orientation of the twist varies in the different mock velocity fields
depending on the angle between the observer's line-of-sight and the
potential.

The kinematic and photometric major axes are aligned in the mock
velocity fields of the spherical cuspy simulations (Figures
\ref{cuspyvfs} and \ref{cuspySNvfs}), and the major and minor
kinematic axes are perpendicular.  The isovelocity contours are
noticeably pinched at the centre of the velocity field.  There is very
little difference between the mock velocity fields of the simulations
with feedback and those without. The deposition of energy due to the
supernovae did not affect the average velocity field in the cold gas
heavily, thus it is still tracing the underlying (dark matter)
potential quite closely.

The mock velocity fields of the cored simulation (Figure
\ref{coredvfs}) are characterised by parallel isovelocity contours at
the centre.  The kinematic major and minor axes are perpendicular, and
the kinematic and photometric major axes are aligned.  The central
regions of these velocity fields are very similar to those of dwarf
and LSB galaxies (e.g. Gentile et al.~2005; K06).

\begin{figure}
\begin{center}
\includegraphics[scale=0.6]{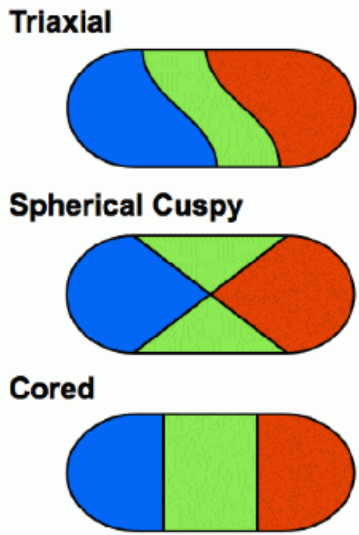}
\caption{Schematic representations of the distinguishing features of
the  triaxial cuspy, spherical cuspy, and spherical cored mock velocity fields.   
\label{schematics}}
\end{center}
\end{figure}

\begin{figure}
\begin{center}
\includegraphics[scale=0.4]{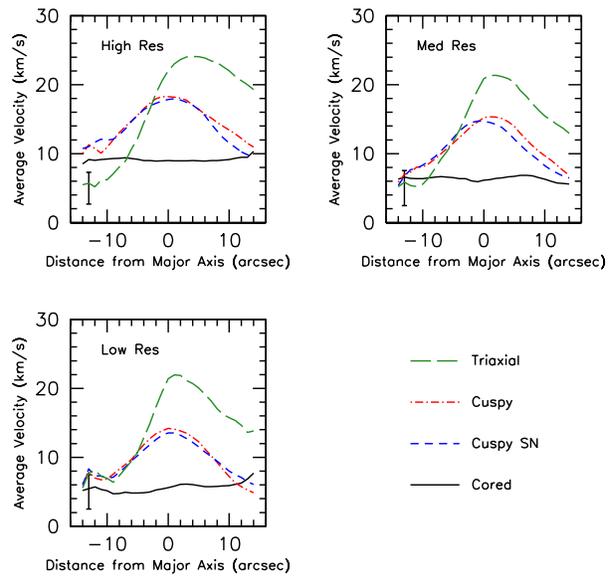}
\caption{Average velocity measured along slits parallel to, and offset
from, the minor axis of the mock velocity fields.  A typical errorbar
is shown in the lower left corner.   \label{slits}}
\end{center}
\end{figure}

We find through visual inspection of the 480 mock velocity fields,
that depending on the underlying halo type, the mock velocity fields
look distinctly different.  We capture the distinguishing features
discussed above in the illustration in Figure \ref{schematics}.  We
can quantify these differences by comparing velocities measured along
slits placed parallel to the minor axis.  In Figure \ref{slits}, we
plot the average of the measured velocities in a
3$\arcsec$~$\times$~28$\arcsec$ slit (the width of a DensePak fiber
and the width of the DensePak array, respectively) parallel to and
offset from the minor axis by $\pm$0.5~kpc.  The velocity is roughly
constant as a function of position along the slit in the cored mock
velocity fields, consistent with the generally parallel isovelocity
contours.  The pinch observed in the spherical cuspy mock velocity
fields translates to a symmetric $\sim$~8~km~s$^{-1}$ rise in velocity
along the slit as the major axis (r~=~0) is approached.  When slits
are placed parallel to the photometric minor axis in the triaxial mock
velocity fields, there is an asymmetric (around r~=~0) overall change
in velocity of $\sim$~16~km~s$^{-1}$ along the slit.  This behavior is
unique to the triaxial case; velocities measured along slits placed at
an angle to the minor axis in the spherical cuspy and cored cases show
a steady change (constant increase or decrease) in velocity along the
entire length of the slit.  It is important to note that the
differences in the central regions of the mock velocity fields are
equally detectable at all three spatial resolutions.  That the
underlying halo type produces a unique and measurable signature in the
observed velocity field is an important result.  It means that the
halo profile can be probed in a model-independent way.

\subsection{Application to LSB galaxy data}

\begin{figure}
\begin{center}
\includegraphics[scale=0.4]{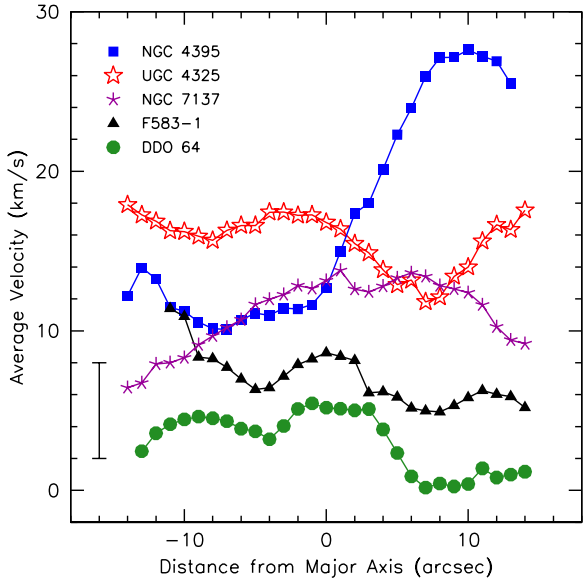}
\caption{Average velocity measured along slits parallel to, and offset
from, the minor axis of galaxies from \citetalias{Kuzio06} and
\citetalias{Kuzio08}.  A typical errorbar is shown in the lower left
corner.    \label{LSBslits}}
\end{center}
\end{figure}

We apply the slit method described in Section \ref{MockVFsection} to
the DensePak velocity fields of the LSB galaxies presented in
\citetalias{Kuzio06} and \citetalias{Kuzio08} to see how well it
compares to the results obtained for the galaxies by fitting halo
models to the rotation curves.  We plot the results for 5 of the
galaxies in Figure \ref{LSBslits}.  \citetalias{Kuzio06} and
\citetalias{Kuzio08} find the cuspy NFW halo to provide a better fit
to the rotation curve of NGC~4395 than the cored isothermal halo; they
also note a misalignment of the minor axis.  In Figure
\ref{LSBslits}, the change in velocity along the slit in NGC~4395
displays the same shape as those along the triaxial slits in Figure
\ref{slits}.  There is an asymmetric (around r~=~0) overall change in
velocity of $\sim$~18~km~s$^{-1}$ along the slit.  Within the errors,
the velocities are roughly constant along the slits for UGC~4325,
DDO~64, and F583-1.  A constant velocity is expected for cored haloes,
and these three galaxies are found to be best-described by cored
isothermal haloes in \citetalias{Kuzio06} and \citetalias{Kuzio08}.
\citetalias{Kuzio08} find the rotation curve of NGC~7137 to be equally
fit by cored and cuspy haloes.  In Figure \ref{LSBslits}, the
velocities along the slit in NGC~7137 display a close to symmetric
(around r~=~0) $\sim$~6~km~s$^{-1}$ change in velocity.  This behavior
is similar to the change in the velocity seen along the spherical
cuspy slits in Figure \ref{slits}.  We find that the slit method can
provide a measure of the underlying halo in real galaxy data that is
in agreement with results based on the traditional method of rotation
curve fitting.

\subsection{Mock rotation curves}

The NEMO \citep{Teuben95} program ROTCUR \citep{Begeman89} is used to
derive a rotation curve from each mock velocity field.  ROTCUR treats
the observed velocity field as an ensemble of tilted rings and fits
for the centre, systemic velocity, inclination, position angle, and
rotation velocity in each ring.  We allow the position angle to change
with radius, but we do not make any explicit attempts to model the
obvious non-circular motions when fitting the triaxial velocity
fields.  For an extensive explanation of ROTCUR and its application to
the DensePak velocity fields, the reader is referred to
\citetalias{Kuzio06} and \citetalias{Kuzio08}.  We also calculate the
rotation curve directly from each of the four galaxy+halo simulations
assuming $V^{2}=GM/r$.

\begin{figure}
\begin{center}
\includegraphics[scale=0.38]{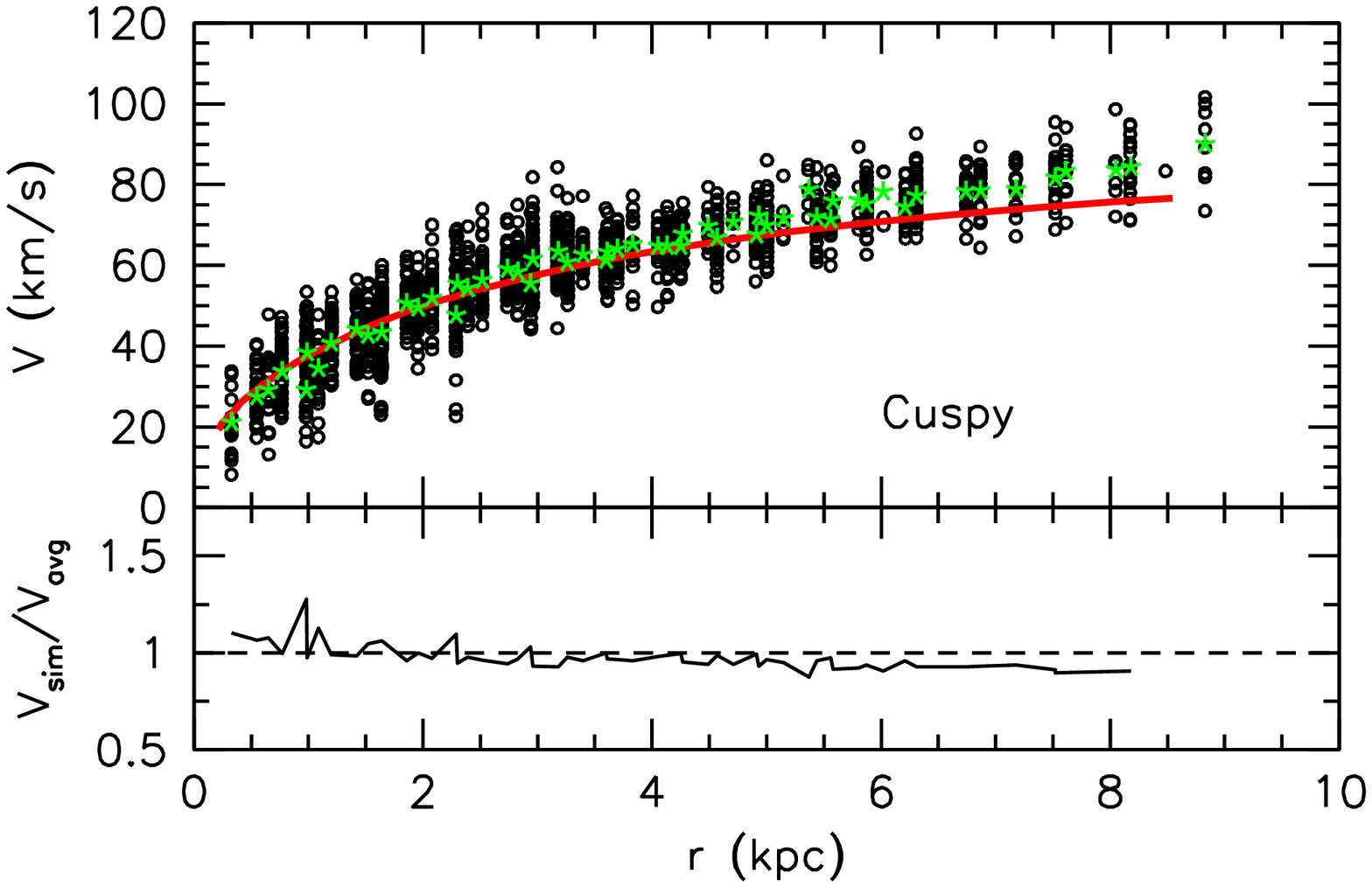}\\
\includegraphics[scale=0.38]{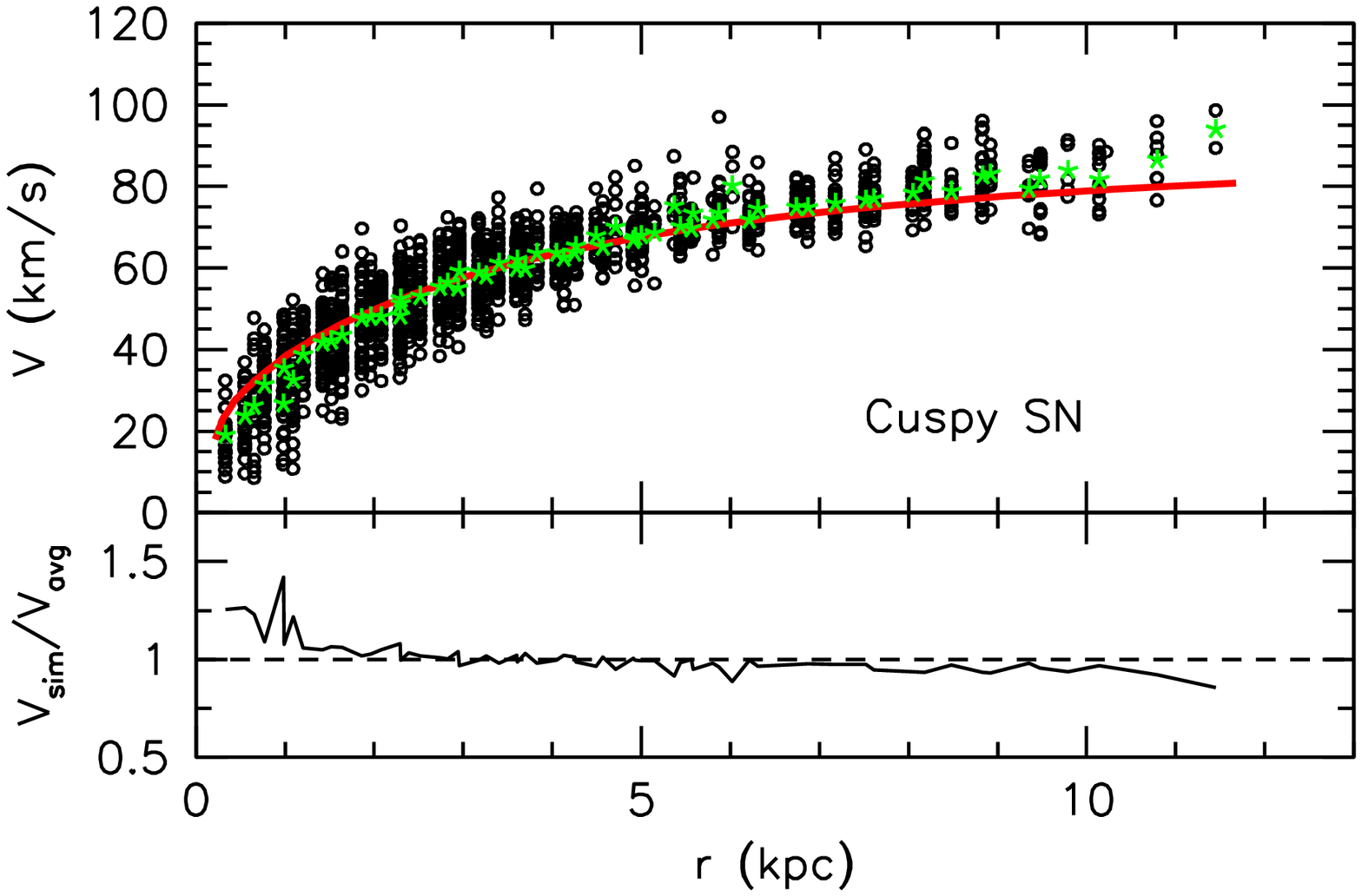}\\
\includegraphics[scale=0.38]{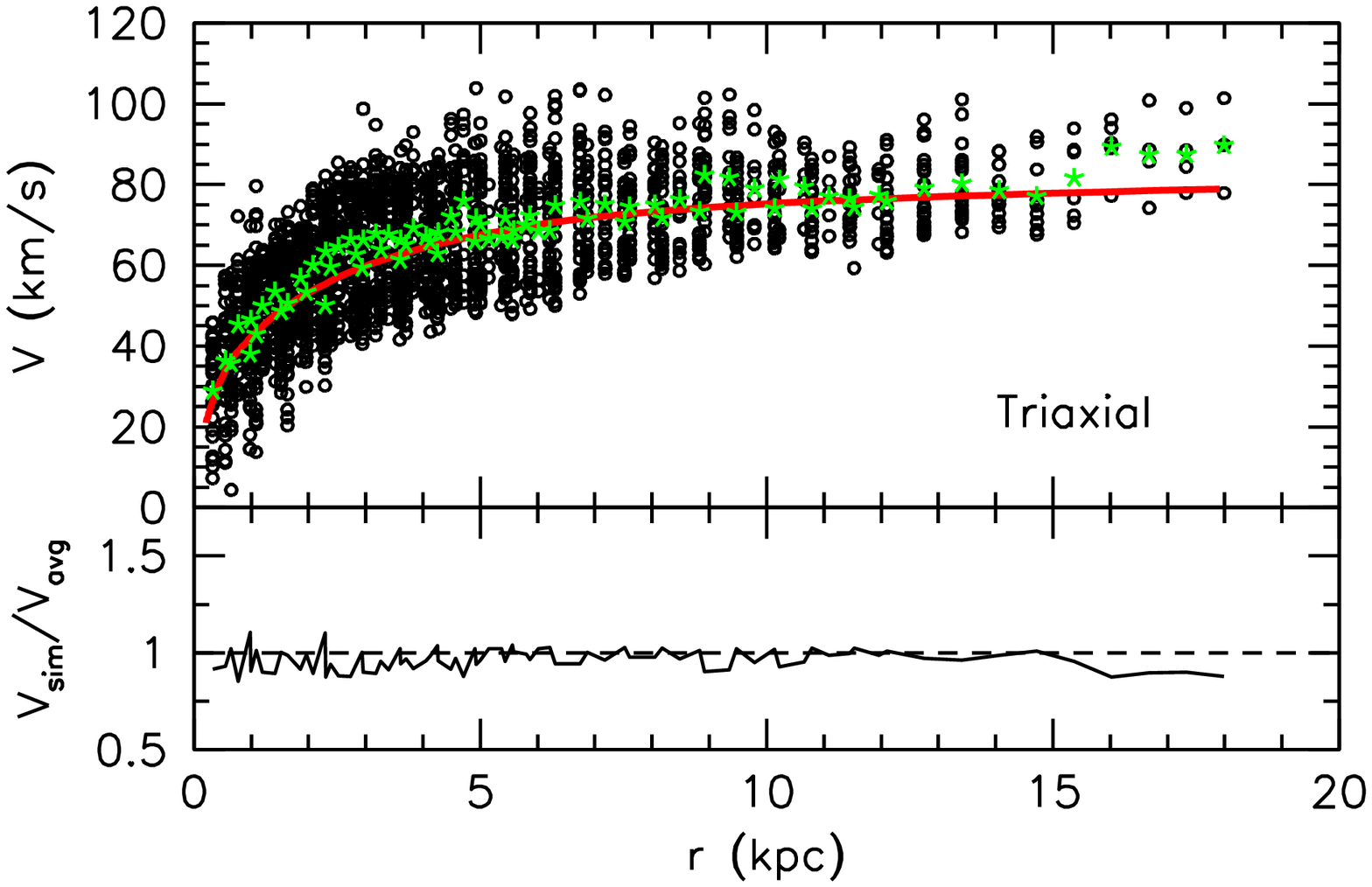}\\
\includegraphics[scale=0.38]{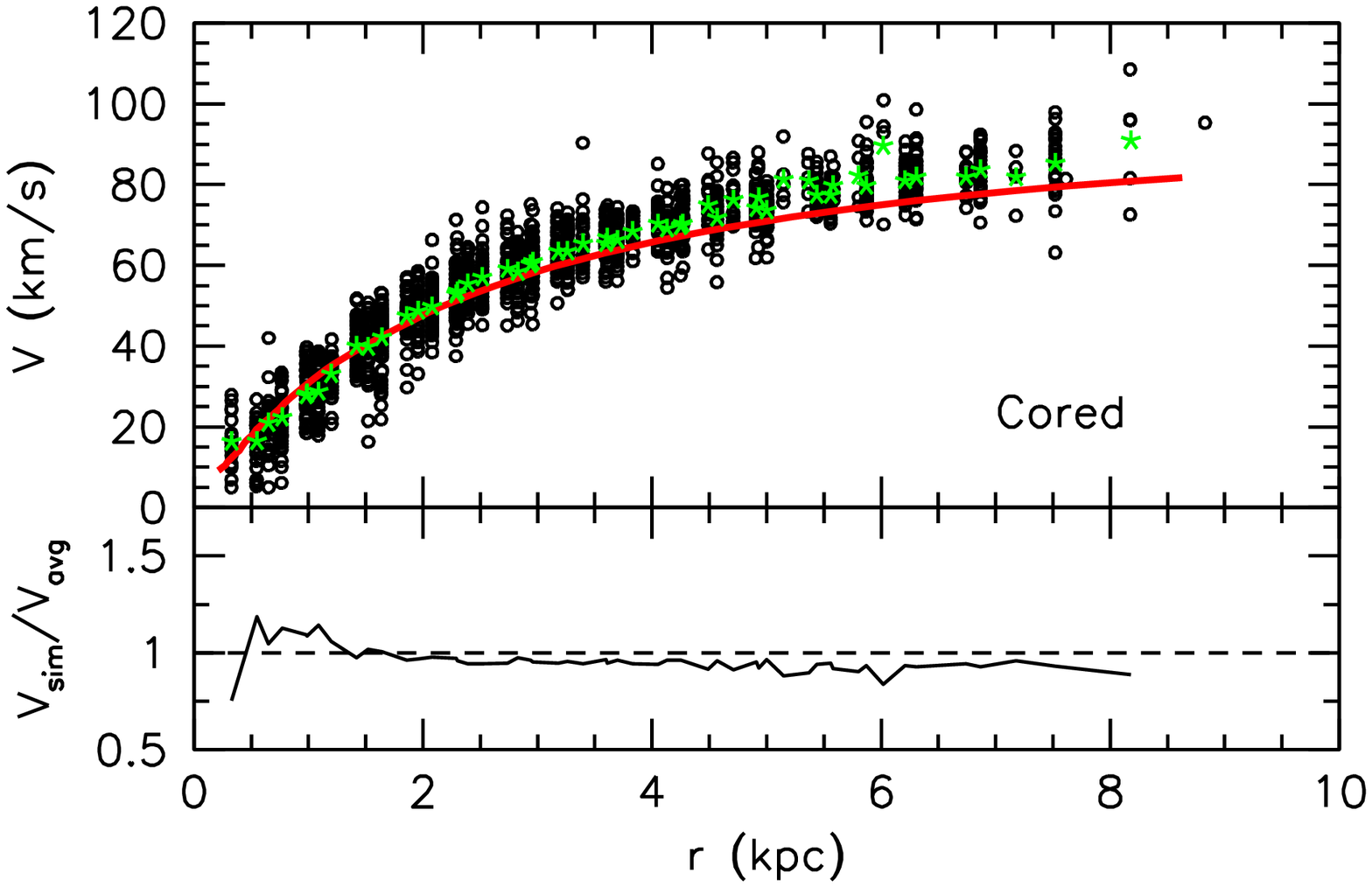}
\caption{Comparison of the recovered mock rotation curves (black open
circles) to the rotation curve derived from the simulation (red line)
using $V^{2}=GM/r$. The green stars are the average mock rotation
curve. The bottom panels plot the ratio of the simulation rotation
curve to the average mock rotation curve.   \label{rcs}}
\end{center}
\end{figure}

In Figure \ref{rcs}, we compare the rotation curves extracted
from each mock velocity field to the rotation curves derived from the
simulations.  Overall, we find the mock rotation curves to trace the
simulation rotation curves well.  The rotation curves extracted from
the triaxial velocity fields show a large spread in velocity at all
radii, not surprising given the dependence on the observer's
line-of-sight with respect to the orientation of the potential.  There
is much less variation in the mock rotation curves extracted from the
spherical cuspy and cored velocity fields.  In addition, the
comparisons of the average mock rotation curves to the rotation curves
derived directly from the simulations are practically
indistinguishable.  In all three cases, the average mock rotation
curve is only $\sim$3 km s$^{-1}$ below the simulation rotation curve
at small radii ( r~$\la$~1~kpc), is equal to the simulation at
intermediate radii, and displays a trend toward slightly higher
average velocities at large radii.

\section{Halo Fits}

 Following the convention of observational studies, we fit each mock rotation curve with both the cuspy NFW halo and the
cored pseudoisothermal halo.     We fit the data under the assumption of negligible
    baryonic mass. The NFW rotation curve is given by
\begin{equation}
V(R) = V_{200}\sqrt{ \frac{\ln(1+cx) - cx/(1 + cx)}{x[\ln(1 + c) - c/(1+c)]}},
\end{equation}
with $x$~=~$R$/$R_{200}$.  Here $R_{200}$ is the radius at which the
density contrast exceeds 200, $V_{200}$ is the circular velocity at
$R_{200}$, and the concentration parameter $c$~=~$R_{200}$/$R_{s} $,
where $R_{s}$ is the characteristic radius of the halo
\citep{Navarro96a,Navarro97}.  The rotation curve of the
pseudoisothermal halo is given by
\begin{equation}
V(R) = \sqrt{4\pi G\rho_{0} R_{C}^{2}\left[1 - \frac{R_{C}}{R}\arctan\left(\frac{R}{R_{C}}\right)\right]} ,
\end{equation}
where $\rho_{0}$ is the central density of the halo and $R_{C}$ is the
core radius of the halo.

\begin{figure}
\begin{center}
\includegraphics[scale=0.35]{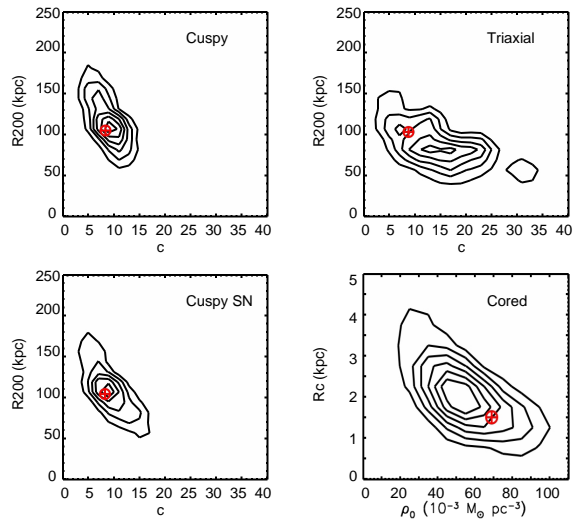}
\caption{Recovered halo parameters from the mock
velocity fields and rotation curves.  The contours represent the
distribution of parameters found with the mock observations and the
red crossed circles indicate the values of the parameters derived
directly from the simulations.  \label{halofits}}
\end{center}
\end{figure}

In Figure \ref{halofits}, we show the distribution of the recovered
halo parameters ($c$, $R_{200}$; $\rho_{0}$, $R_{c}$) from the fits to the mock data (In the following,  $R_{200}$, $\rho_{0}$ and $R_{c}$ are given in units of kpc, $10^{-3} \Msun$ pc$^{-3}$ and kpc, respectively).  We find that the
parameter values obtained from the fits to the mock rotation curves
are in agreement with those derived directly from the simulations
(Table \ref{parameters}).  The parameters of the best-fitting NFW halo
to the spherical cuspy data are $c = 9.4\pm1.8$ and $R_{200} =
107\pm10$; for the spherical cuspy with feedback data, the values are
$c = 8.5\pm3.0$ and $R_{200} = 113\pm20$.  There is more variation in
the halo fits to the triaxial  cuspy data.  The best-fitting halo parameters
are $c = 11.0\pm4.5$ and $R_{200} = 87\pm10$.  The parameters of the
best-fitting isothermal halo to the spherical cored data are $R_{c} =
2.0\pm0.5$ and $\rho_{0} = 50\pm5$.

 Fitting a pseudoisothermal sphere to the cuspy data results in 
  dense, small cores ($\rho_{0}\ga$80; $R_{c}\la$1.0); 
  similarly, when fitting cuspy haloes to the cored data, the data want $c <
  1.0$ and high $R_{200}$. This is not surprising since a profile with a tiny core and
  high central density resembles a cuspy profile and only with a
  very low concentration can one stretch an NFW curve to fit the
  rotation curve of a cored halo, see
  \citet{deBlok10}. However, given the information from the velocity fields, 
  one can easily break this ambiguity and identify the underlying
  halo shape even before fitting a profile to it. Additionally, the
  mass-concentration relation of CDM \citep{Maccio08} would predict
  $c < 1.0$ halos to have rotation velocities many times higher
  than the galaxy itself.

\begin{figure}
\begin{center}
\includegraphics[scale=0.38]{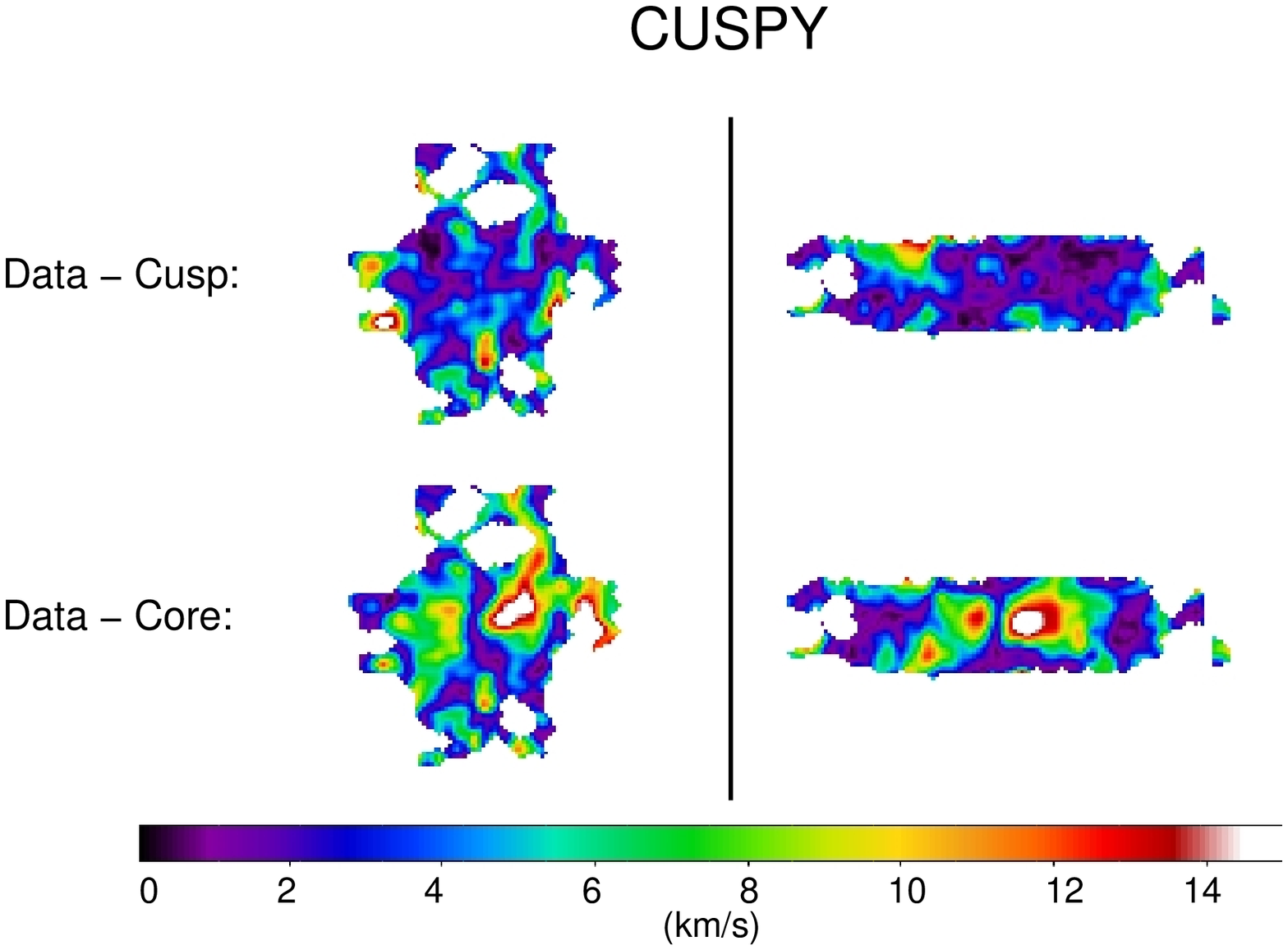}\\
\vspace{15pt}
\includegraphics[scale=0.38]{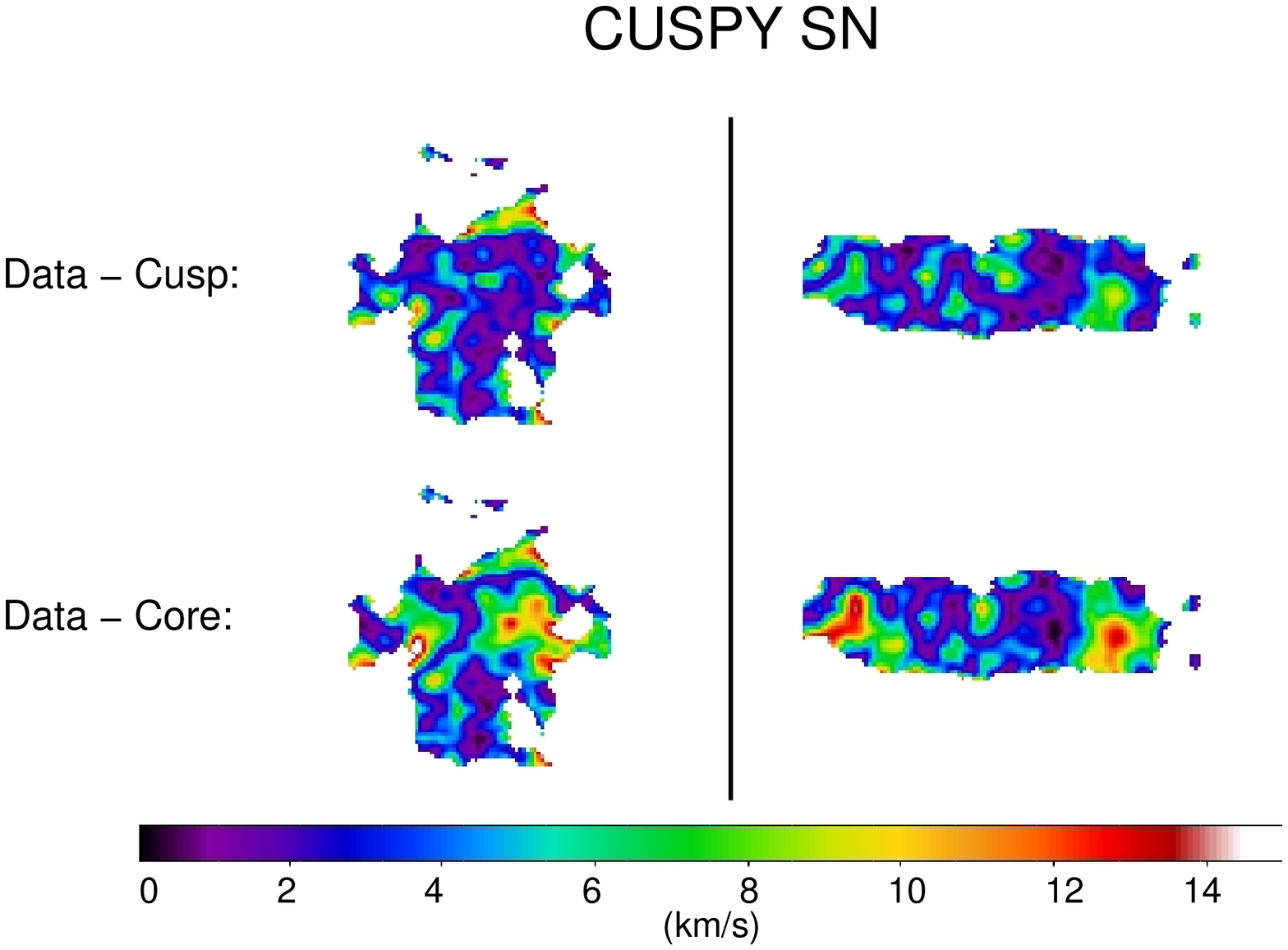}\\
\vspace{15pt}
\includegraphics[scale=0.38]{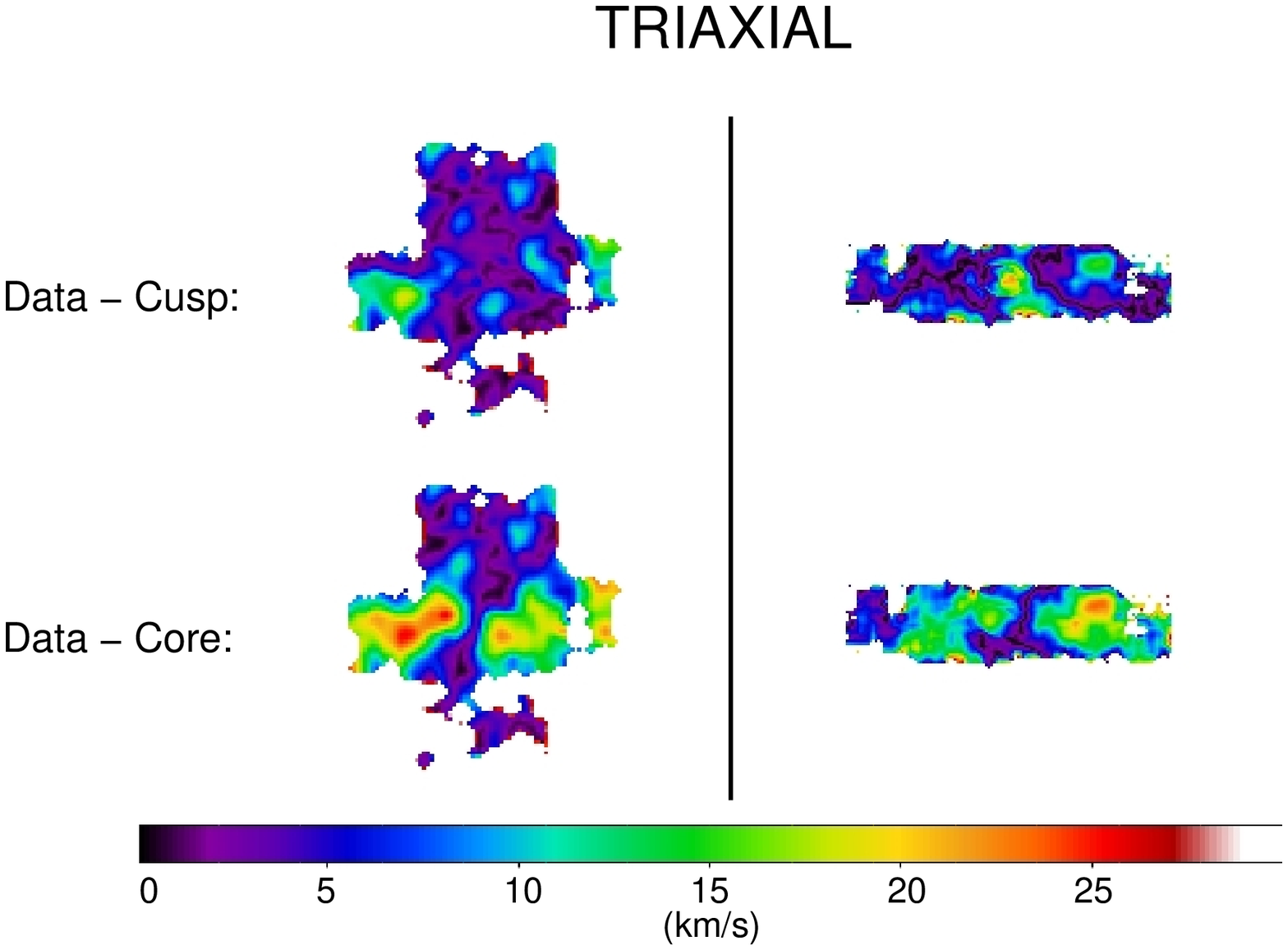}\\
\vspace{15pt}
\includegraphics[scale=0.38]{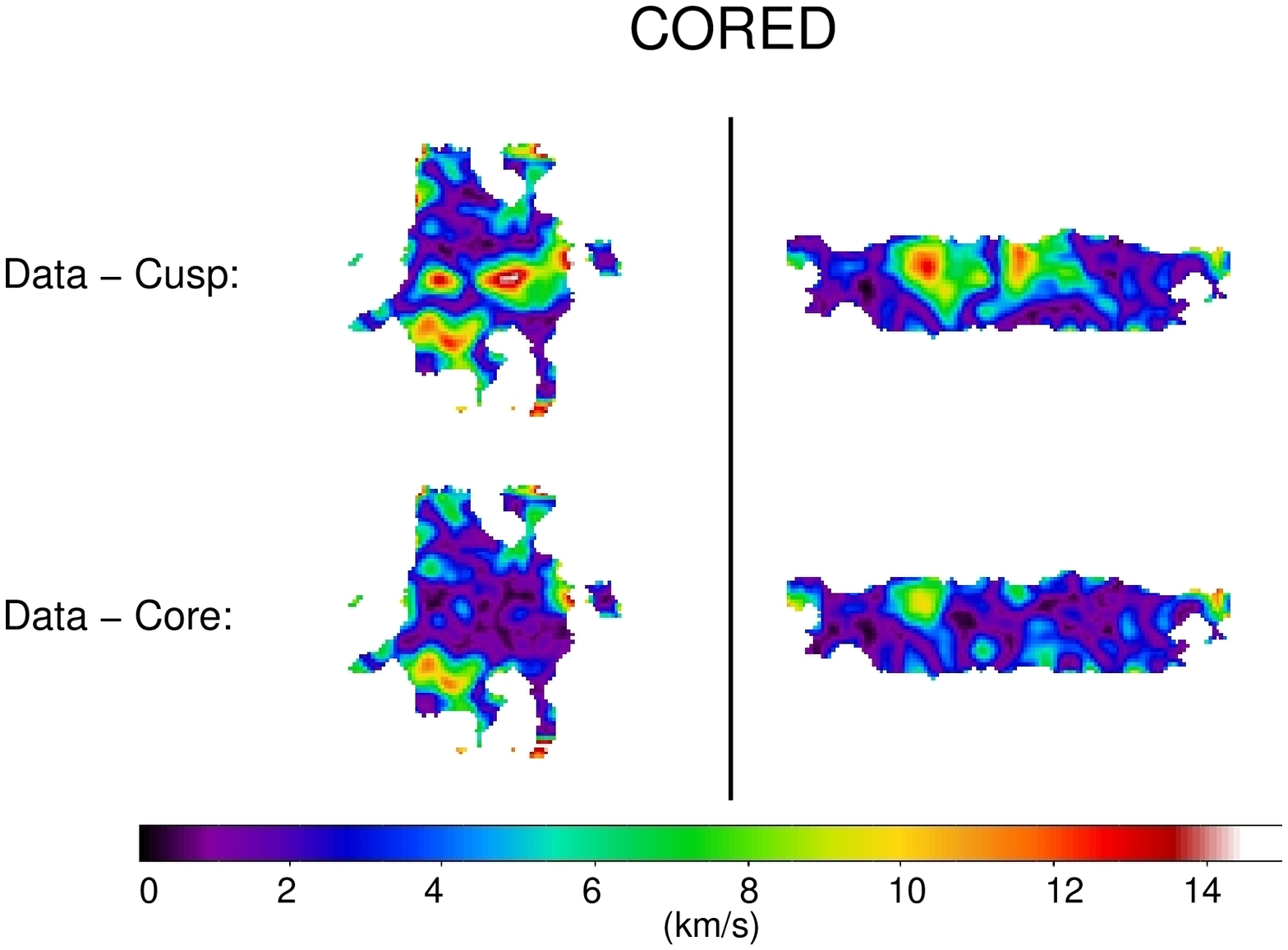}
\caption{Sample residual velocity fields after subtracting the
best-fitting cuspy and cored halo models from two different observed
mock velocity fields (Realistic 1 case on the left, Realistic 2
case on the right.)   \label{residuals}}
\end{center}
\end{figure}

In Figure \ref{residuals} are representative examples of residual mock
velocity fields showing the differences between the mock data and the
best-fitting cuspy and cored halo models.  For the cored mock velocity
fields, we display the residuals obtained when subtracting the
best-fitting spherical cuspy halo model; the residuals obtained when
comparing the cored data to the best-fitting spherical cuspy with
feedback halo model are comparable.  The residuals are small,
typically less than $\sim$~4~km~s$^{-1}$, when cusps are fit to cusps
and cores are fit to cores.  When a core is fit to the cuspy data or a
cusp is fit to the cored data, the residuals can become quite large.
Residuals as high as $\sim$~15~km~s$^{-1}$ and $\sim$~30~km~s$^{-1}$
are obtained when cores are fit to the spherical and triaxial cuspy
data, respectively; residuals on the order of 15~km~s$^{-1}$ are
obtained when cusps are fit to the cored data.

 We also calculate the inner slope of the mass density profile by
converting the mock rotation curves using
\begin{equation}
\rho(R) = \frac{1}{4\pi G}\left[2\frac{V}{R}\frac{\partial V}{\partial R}+\left(\frac{V}{R}\right)^{2}\right],
\end{equation}
where $G$ is the gravitational constant and $V$ is the observed rotation velocity at radius $R$ \citep{deBlok01}.

\begin{figure}
\begin{center}
\includegraphics[scale=0.35]{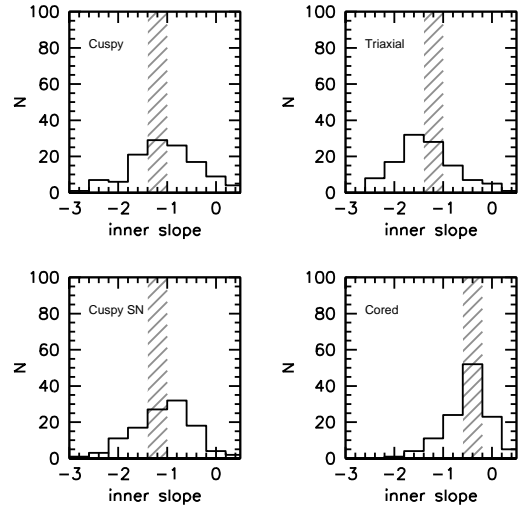}
\caption{Distribution of the asymptotic inner slope of the dark matter density profiles derived from the mock rotation curve data.  The gray shaded regions indicate the inner slope derived directly from the dark matter densities of the simulations. \label{innerslope}}
\end{center}
\end{figure}

In Figure \ref{innerslope}, we show the distribution of the asymptotic inner slope of the dark matter density profiles derived from the mock rotation curve data.  The gray shaded regions indicate the inner slopes derived directly from the dark matter densities of the simulations.  We find the cuspy halos to generally have inner slopes of $-1$ or steeper and the inner slope of the cored halo to be more shallow, around a value of $\sim-0.4$.  The inner slopes derived directly from the simulations are $-1.3$, $-1.4$, $-1.4$, and $-0.3$ for the spherical cuspy, spherical cuspy with feedback, triaxial cuspy, and spherical cored halos, respectively.  The corresponding inner slopes determined from the mock data are $-1.0\pm0.7$, $-1.0\pm0.8$, $-1.4\pm0.7$, and $-0.5\pm0.4$.

\section{Discussion and Conclusions}

We have presented mock velocity fields, rotation curves, and halo fits
for simulated LSB galaxies formed in spherical and triaxial cuspy dark
matter haloes and spherical cored dark matter haloes.  The mock
velocity fields span a range of data quality representing ideal to
realistic observations.  The main findings of this work are:
\begin{itemize}
\item The underlying halo type produces a unique signature in the
  velocity field. This can be used to constrain the shape of the dark
  matter profile (spherical or triaxial cuspy, or spherical cored)
  without fitting an analytic density profile to it.

\item Cored and cuspy haloes can also be distinguished clearly by
  deriving their asymptotic inner slopes from the rotation curve data.

\item Given at least one of the above information one can then
  successfully recover the underlying halo parameters from the
  rotation curve using the appropriate analytic form for the density
  profile (NFW or pseudoisothermal sphere).
\end{itemize}
\textit{This means that if an LSB galaxy were in a cuspy halo, the
  cusp would be observable in the data.}  Given these results, we find
it difficult to mistake cuspy haloes, spherical or triaxial, for cored
haloes. The observed cores in dark matter-dominated galaxies are true
discrepancies from the predictions of (dark matter-only)
$\Lambda$CDM simulations.  Systematic effects, non-circular motions,
and halo triaxiality cannot explain the observed differences.

Baryonic processes more effective than those we have modeled or
entirely different dark matter models may be necessary to explain the
observed cores.  Feedback from star formation could transform the
initial cusps into cores.  If LSB galaxies form with a massive dark
matter halo already in place and with the same low baryon fraction
that we observe today, then feedback from star formation of the type
we have modeled would not be enough to change a cusp into a core.  In
this scenario, the dark matter halos in which LSB galaxies reside
cannot be cuspy CDM halos.  If, however, feedback processes from star
formation can affect dark matter haloes during the early stages of
their formation, it is possible that cusps can be transformed into
cores.  This has been demonstrated by \citet{Governato10} (see also
\citet{Oh10}) using a set of fully cosmological simulations of dwarf
galaxy formation ($M_{vir} \sim 3 \times 10^{10} \Msun$) where strong
outflows from supernovae remove low-angular-momentum gas and decrease
the dark matter density to less than half in the central part 
 \citep[see also][]{Navarro96b,Read05,Mashchenko08}. 
The
question in this scenario then becomes one of detecting the blown-out
baryons.  It also remains to be seen if this kind of mechanism can act
efficiently on higher mass scales, given the deeper potential wells.

Alternatively, the observed cores may need to be addressed by a
different dark matter model/particle that naturally produces cored
haloes.  Recent results, however, have found that self-interactions
and warm dark matter particle properties cannot be responsible for the
cores observed in dark matter-dominated galaxies
\citep[e.g.][]{Kuzio10,Villaescusa10}.
\section*{Acknowledgments}

The work of R.~K.~D.~was supported by NSF Astronomy \& Astrophysics
Postdoctoral Fellowship grant AST 0702496. T.~K.~acknowledges
financial support from the Swiss National Science Foundation (SNF).
It is a pleasure to thank James Wadsley, Joachim Stadel and Tom Quinn
for making \textsc{Gasoline} available to us. We acknowledge useful
and stimulating discussions with James Bullock, Fabio Governato and
Stacy McGaugh.


\end{document}